# Potential Networking Applications of Global Positioning Systems (GPS)[1]

Gopal Dommety, Raj Jain
Department of Computer and Information Science
The Ohio State University
Columbus, OH 43210-1277
Contact: Jain@ACM.Org

## Abstract

Global Positioning System (GPS) Technology allows precise determination of location, velocity, direction, and time. The price of GPS receivers is falling rapidly and the applications are growing. PCMCIA receivers that can be connected to any notebook personal computers are available for $300-400 to end consumers.

The main goal of this study was to survey current applications of GPS to distributed systems and networks. While GPS is appearing in the computer magazines very often and while many computer companies have announced GPS related efforts, most such efforts are in providing navigational guidance to drivers. Digitized city maps along with a GPS sensor on a mobile computer provide directions to drivers. A number of consortiums have been formed by companies such as IBM, Apple, Toshiba, Sony, and others.

Currently, applications of GPS for distributed computation and networking are limited to measuring delays in Wolter and Golderman's DA-30 network analyzers and in clock synchronization in synchronous optical networks (SONET) used in telecommunication networks.

We have identified twenty new applications of GPS for distributed computing and networking. These applications include circuit switching using synchronized clocks, synchronous slotted systems, clock synchronization in distributed systems, database synchronization, connectionless real-time communication, one-way delay, delay based routing, time to live, resource location, location adaptive protocols, home vs office vs car, electronic fence, handoffs in wireless networks, prescheduled handoffs based on velocity and direction, adaptive transmission power control algorithms, directional antennas, temporary cell partitioning for congestion avoidance, peer-to-peer routing with limited range receivers, email delivery based on geographic location, distributed robot control and navigation, and equipment location marking for maintenance crew.

There two main obstacles to applications of GPS. First is that GPS antennas must point to open sky. They cannot be used directly inside a building. Unless this is avoided by new antenna designs or by rebroadcasting GPS data inside a building, the use of GPS techniques is limited to improving the performance of distributed systems rather than proper operation. This is similar to cache memories in computer systems. Systems can operate without cache but perform better with cache. The second obstacle is the psudo-random noise introduced in the GPS signals by the defense department to disallow other governments from using the full precision of GPS. This is called selective availability. Fortunately, this can be easily overcome by differential GPS techniques.

---

[1] We would like to thank 3COM Corporation for partially sponsoring this project.

Detailed lists of GPS products, current applications, addresses of manufacturers, and sources for further information are included in this report.

# Contents

# 1 Introduction

Global Positioning Systems (GPS) are space-based radio positioning systems that provide time and three-dimensional position and velocity information to suitably equipped users anywhere on or near the surface of the earth (and sometimes off the earth). Concept of satellite navigation was first conceived after the launch of Sputnik 1 in 1957 when scientists realized that by measuring the frequency shifts in the small bleeps emanating from this first space vehicle it was possible to locate a point on the earth's surface. The NAVSTAR system, operated by the US Department of Defense, is the first such system widely available to civilian users. The Russian system, GLONASS, is similar in operation and may prove complimentary to the NAVSTAR system. Current GPS systems enable users to determine their three dimensional differential position, velocity and time.

These systems will change our lives in many ways. By combining GPS with current and future computer mapping techniques, we will be better able to identify and manage our natural resources. Intelligent vehicle location and navigation systems will let us avoid congested freeways and find more efficient routes to our destinations, saving millions of dollars in gasoline and tons of air pollution. Travel aboard ships and aircraft will be safer in all weather conditions. Businesses with large amounts of outside plant (railroads, utilities) will be able to manage their resources more efficiently, reducing consumer costs.

# 2 Principles of GPS

The GPS satellites broadcast signals, which contain information enough to compute the distance ($x$) of the location from the satellite. Once the distance is known, the point in question can be anywhere on a circle of radius $x$ (in 2-D and sphere in 3-D) with the satellite as the center. By knowing the distance from another satellite, the possible positions of the location are narrowed down to two points (Two intersecting circles have two points in common). To distinguish between the two locations, distance from another satellite can be computed or the ridiculous answer can be eliminated. This assumes a precise clock at the GPS receiver. Inaccuracies in the clock can lead to erroneous results if using two satellites or inconsistent equations if using three. To over come this limitation, the distance from three satellites are computed and algebraic laws are applied to achieve the correct position. Accurate 3-D measurements require four satellites. To achieve 3-D real time measurements, the receivers need at least four channels. Receivers equipped with only one channel have to measure the distance sequentially from multiple satellites and can take 2 to 30 seconds. For further details on the working of GPS, see [15].

# 3 NAVSTAR Global Positioning System

NAVSTAR is a constellation of 24 satellites (3 of which are spare) orbiting the earth at a height of 10,900 nautical miles with an orbital period of 12 hours and a planned life-span of 7.5 years. These orbits are inclined 55 degrees to the equatorial plane. In any event, the constellation will provide a minimum of four satellites in good geometric positions. Up to 10 GPS satellites are normally seen assuming a 10 degree elevation. The satellites are equipped with several cesium clocks, which provide high timing accuracies. The satellites are in synch with each other. Positional accuracy of 100 meters, timing accuracy of 300 nanoseconds and frequency accuracies of a few parts in $10^{12}$ can be obtained. These accuracies are achievable with the selective availability (artificial degradation) in effect.

Each satellite transmits on two L band frequencies, L1 (1575.42 MHz) and L2 (1227.6 MHz). L1 carries a precise (P) code and a coarse/acquisition (C/A) code. L2 carries the P code. The P code is normally encrypted so that only the C/A code is available to civilian users; however, some information can be derived from the P code. Incidentally, when P code is encrypted it is known as Y code. A navigation data message is superimposed on these codes. The same navigation data message is carried on both frequencies.

There have been three distinct groups of NAVSTAR satellites so far. The groups are designated as blocks. The block I satellites were intended for system testing. The block II satellites were the first fully functional satellites, including cesium atomic clocks for timing as well as the ability to implement selective availability. The latest satellites, the block IIR versions, include autonomous navigation.

Satellite operating parameters such as navigation data errors, signal availability/anti-spoof failures, and certain types of satellite clock failures are monitored internally within the satellite. If such internal failures are detected, users are notified within six seconds. Other failures may take from 15 minutes to several hours to rectify. Some receivers compute an estimated figure of merit, which indicates the confidence level of the position information. The development of integrity capabilities to meet safety requirements of various applications is underway.

Each satellite has two identifying numbers. First is the NAVSTAR number which identifies the specific satellite hardware. Second is the space vehicle (SV) number. This number is assigned in order of launch. The third is the pseudo-random noise code number (PRN). This is a unique integer number which is used to code the signal from that satellite. Some receivers identify the satellites that they are listening to by SV, others by PRN. The NAVSTAR system is sponsored by the Department of Defense to provide two sets of services SPS and PPS - explained next.

## 3.1 Standard Positioning Service (SPS)

SPS provides standard level of positioning and timing accuracy that is available, without qualification or restrictions, to any user continuously on a worldwide basis. Most receivers are capable of receiving and using the SPS signal. The SPS accuracy is intentionally degraded by the DOD by the use of *Selective Availability* based on US security interests. This service provides 100 meters horizontal accuracy, 156 meter vertical accuracy and 167 nanoseconds time accuracy. (The signals providing standard positioning service are inherently capable of greater accuracy than this).

## 3.2   Precise Positioning Service (PPS)

PPS is the most accurate positioning, velocity, and timing information continuously available, worldwide, from the basic GPS. PPS signals can only be accessed by authorized users with cryptographic equipment and keys and specially equipped receivers. US and Allied military, certain US Government agencies, and selected civil users specifically approved by the US Government, can use PPS. The accuracies achievable using PPS are 17.8 meter horizontal accuracy, 27.7 meters vertical accuracy and 100 nanoseconds time accuracy

## 3.3   Selective Availability (SA)

SA is the intentional degradation of the SPS signals by a time varying bias by the DOD to limit accuracy for non-US military and government users. The potential accuracy of the C/A code is around 30 meters, but it is reduced to 100 meters. The SA bias on each satellite signal is different, and so the resulting position solution is a function of the combined SA bias from each SV used in the navigation solution. Differential corrections must be updated at a rate less than the correlation time of SA (and other bias errors).

# 4   Differential GPS

Differential GPS (DGPS) is a method of eliminating errors in a GPS receiver to make the output more accurate. This process is based on the principle that most of the errors seen by GPS receivers in a local area will be common errors. These common errors are caused by factors such as clock deviation, selective availability (explained in Section 3.3), drift from predicted orbits, multipath error, internal receiver noise and changing radio propagation conditions in the ionosphere. If a GPS receiver is placed at location for which the coordinates are known and accepted, the difference between the known coordinates and the GPS-calculated coordinates is the error. This receiver is often called a "base station."

The error, which the base station has determined, can be applied to other GPS receivers (called "rovers"). Since the sources of the error are continuously changing, it is necessary to match the error correction data from the base station very closely in time to the rover data. One way of doing this is to record the data at the base station and at the rover. The data sets can be processed together at a later time. This is called post processing and is very common for surveying applications. The other way is to transmit the data from the base station to the rover. The error calculation is made in the rover in real time. This process is called real-time DGPS.

Differential GPS offers accuracies of a few meters (better than the military's precise positioning service). It also facilitates detection of erroneous signals from the satellites. Currently commercial services exist for getting real time differential data. Development of DGPS services is one of the options considered by the Joint DOD/DOT task force on the issue of GPS-standard position service accuracy.

# 5 Other Radio Navigation Systems

## 5.1 GLONASS

The GLONASS constellation is composed of 24 satellites, eight in each of three- orbital planes. The satellites operate in circular 19,100 km orbits at an inclination angle of 64.8 degrees and with a 11-hour, 15 minute period. Each satellite transmits on two L frequency groups. The L1 group is centered on 1609 MHz while the L2 group is centered on 1251 MHz. Each satellite transmits on a unique pair of frequencies. The GLONASS signals carry both a precise (P) code and a coarse/acquisition (C/A) code. The P code is encrypted for military use while the C/A code is available for civilian use.

## 5.2 LOng-RAnge Navigation (LORAN-C)

Loran was one of the earliest and the most successful systems for ground-based radio navigation. Two versions are currently in operation: Loran C, which serves civilian users and Loran D, which serves the military. These are medium to long range, low frequency time-difference measurement systems. A master and usually up to four secondary transmitting stations put out a set of radio pulses centered on 100 kHz, in a precisely timed sequence. The receiver measures the difference in arrival time between these transmissions from different stations and estimates the position. Loran C transmissions can be worked out to ranges over 1500 kilometers from master stations, providing accuracies of 100 to 500 meters. Shorter range accuracies of better than 30 meters are also available.

## 5.3 TRANSIT

Transit was the first operational satellite navigation system. Developed by Johns Hopkins Applied Physics Laboratory, the system was intended as an aid to submarine navigation.

The Transit system allowed users to determine position by measuring the Doppler shift of a radio signal transmitted by the satellite. Users were able to calculate position to within a few hundred meters as long as they knew their altitudes and the satellite ephemeris.

The system has several drawbacks. First, the system is inherently two dimensional. Second, the velocity of the user must be taken into account. Third, mutual interference between the satellites restricted the total number of satellites to five. Thus, satellites would only be visible for limited periods of time. These drawbacks pretty much eliminate aviation applications and severely limit land-based applications.

## 5.4 Timation

Developed in 1972 by the Naval Research Laboratory (NRL), Timation satellites were intended to provide time and frequency information. The original satellite flew with stable quartz crystal oscillators. Later models flew with the first space-borne atomic clocks. The third satellite acted as a GPS technology demonstrator.

## 5.5 Low-altitude Satellites

Experiments are also being proposed to place GPS receivers on low altitude satellites with various other sensing devices such as laser altimeters and synthetic aperture radar. This could greatly reduce the cost of determining position.

# 6 Accurate Time using GPS

Internally in the NAVSTAR system, time is kept as GPS time. GPS Time began on January 6, 1980 and is referenced in GPS weeks and seconds. GPS time is a composite time composed of the times of all available satellite and monitor station clocks. It is monitored by the GPS Operational Control System and by the US Naval Observatory and is steered to keep it within 1 microsecond of UTC. However, leap seconds are not inserted so GPS time lags behind UTC. The exact difference is provided as constants in the GPS navigation message. By using the information given in the NAVSTAR message, the user can infer UTC time via GPS with a precision of better than 340 nanoseconds (95% probability) using the standard positioning service and 100 nanoseconds using the precise positioning service.

GPS can be used to determine accurate time globally. The GPS time system has a constant offset of 19 seconds with the International Atomic Time (IAT). GPS belongs to the dynamic system achieved by the atomic time scales.

Inexpensive GPS receivers operating at known positions provide a timing accuracy of about 0.1 microsecond with only one satellite in view. With more sophisticated techniques, one can globally synchronize clocks precisely. Presently achievable accuracy of time via GPS is some tens of nanoseconds, however one nanosecond is considered possible.

Ultra precise time transfer is possible (few nanoseconds) but requires advanced preparation, coordination of the two sites and tracking of specific satellites during specific time periods.

## 6.1 Time and Frequency Alternatives

In the United States, National Institute of Standards and Technology (NIST) and US Naval Observatory (USNO) provide services to calibrate frequency references to the internationally accepted definition. NIST provides WWV and WWVH radio broadcast stations (accurate to one millisecond of UT1) and WWWVB broadcast stations (2 to 3 parts in $10^{11}$).

USNO provides Loran-C (LOng RAnge Navigation) a land based radio navigation system. Loran-C transmissions are also used for time dissemination and frequency reference purposes. Typically frequency accuracies of 1 part in $10^{12}$ and time accuracies of better than one microsecond can be achieved.

Both USNO and NIST provide telephone voice messages (accuracy 30 milliseconds), computer modem time transfer (several milliseconds) and remote synchronization of time bases (1 part in $10^9$).

# 7 GPS Data Format Standards

If the user desires improved accuracy at the time the equipment is being used, real-time processing must be employed. For real-time processing, special formats are employed. There are two predominant formats currently being employed.

NMEA-0183 is a data format standard commonly employed for communications between ship-borne navigation electronics. GPS receivers output this format but do not accept it.

The second format, RTCM-104 is an attempt by the Radio Technical Commission for Maritime (RTCM) Services to standardize DGPS operation. The standard is the result of a request by the Institute of Navigation to the RTCM to develop recommendations for DGPS transmission. The RTCM formed Special Committee 104 titled "Differential NAVSTAR GPS Service." Version 2 of this service is used by many beacon systems (including the US Coast Guard system). Version 2.1 includes additional information for the transfer of real-time kinematic data.

# 8 Time Synchronization Techniques

Categories of time synchronization techniques currently employed are absolute time synchronization, clock fly-overs, common view mode, and multi-satellite common view mode.

With absolute time synchronization a special time synchronization receiver picks up the signals from a single GPS satellite as it travels overhead. With this simple technique accuracies of about 100 nanoseconds (with S/A off) and 300 nanoseconds (with S/A on) are feasible.

When clock fly-overs are used, the satellite swings up over the two sites, one after the other. As it passes over each site, careful clock synchronization operations are performed to determine for each ground based clock relative to the satellite. Under typical conditions this mode yields a clock synchronization error of around 50 nanoseconds.

The common view mode can be implemented when two distant sites have direct line-of-site access to the same GPS satellite at the same time. This mode yields synchronization errors of 10 nanoseconds or less. The multi-satellite common view mode involves four or more satellites that are being observed simultaneously from the two different clock sites. Experts believe that they can achieve synchronization errors as small as 1 nanoseconds.

# 9 Current General Applications of GPS

GPS is coming in wide use. With increasing sales volume, the costs are coming down and new applications are being reported in the news everyday. However, very few applications of GPS in distributed systems and networking have been reported. Before discussing them, it is useful to take a brief look at other applications. These can provide us new ideas about computer applications.

## 9.1 Frequency Counters

GPS receivers can be optimized for frequency and time applications. Accurate frequency counters, time interval counters, frequency calibrators and phase comparators can be built using the GPS technology. Accurate frequency and timing offset measurements can be made.

The GPS clock module of the Stellar GPS Corporation (now Absolute Time Corporation) is a L1 C/A GPS receiver optimized for frequency and time applications. The GPS clock provides the user with a very stable 10 MHz reference frequency. In the GPS clock, one pulse per second signal exhibits an RMS pulse-to-pulse jitter of 1 nanosecond rather than 20 to 60 nanoseconds observed on 1 pps outputs from time based GPS receivers. It has the capability to do frequency/timing offset measurements (timing resolution is 5 nanoseconds and frequency resolution of one in $10^{12}$) [19].

## 9.2 Intelligent Vehicle Highway Systems (IVHS)

IVHS will combine GPS technology with communications, controls, navigation and information systems to improve highway safety, ease traffic congestion, and reduce harmful environmental effects.

## 9.3 Car Navigation Systems

A car navigation system uses a specialized computer that use the signals from GPS satellites to track the driver's progress on a digital map. It may provides services like the shortest route, etc. This is the most publicized application of GPS.

## 9.4 Geographic Information Systems (GIS)

GPS technology is used extensively in geographic information systems. These systems combine cellular data networks for communication, GPS for vehicle location, and geographic information system tools for mapping display.

Use of GPS in GIS will permit state and local governments to more efficiently coordinate roadway maintenance and construction in rural areas, provide efficient ways of maintaining roadway databases, and maintain accident inventories.

## 9.5 Emergency Systems

GPS technology is being used to develop emergency messaging products. With the aid of a wireless communication link, the emergency system communicates the GPS derived position information and the specifics of the situation to the base station. GPS based car alarms with GPS to locate stolen cars are being considered by the automobile industry. Several other emergency products exist currently.

## 9.6 Aviation

GPS technology is being applied in aircraft safety systems, air traffic control system, and zero visibility landing. GPS technology can be used to plot aircraft altitude to a pitch of one-tenth of one degree. In future, pilots will do more monitoring, while computers will issue air traffic instructions. This is expected to reduce the number of people required in the control tower and cockpit.

## 9.7 Astronomical Telescope Pointing

Astronomers attempt to observe the occultation of stars by asteroids. This allows them to determine the sizes and shapes of asteroids. To observe these events, astronomers have to use mobile telescopes placed in the predicted path of the shadow. The Odetics GPStar 325 GPS receiver feeds coordinates to a MicroVAX II computer that then aims the telescope at the asteroid. Timing signals from the GPS receiver are used to discipline a crystal oscillator that is used to time the event. [11]

## 9.8 Atmospheric Sounding using GPS Signals

Radio occultation techniques allow researchers to make observations about planetary atmospheres. During the Mariner and Voyager missions, spacecraft trajectories were planned so that radio signals from the spacecraft would transect a planets atmosphere. Earth-based receivers were able to detect phase changes in the radio signals based on the refractivity of the atmosphere.

By placing a GPS receiver in low earth orbit, hundred occultations of GPS signals could be observed every day. Information obtained in this fashion could include atmospheric refractivity, density, pressure temperature and humidity.

## 9.9 Tracking of Wild Animals

Animals are equipped with GPS receivers and with wireless transmitters. The GPS determined position is transmitted to the control station. This information is used to track animals and for studying their nomadic patterns.

## 9.10 GPS Aides for the Blind

GPS determined position can be used to locate a users position on a digitized geographical maps. Real time GPS along with digitized maps and possibly audio capability can provide useful navigational capabilities to the blind.

## 9.11 Recorded Position Information

The recorded GPS position information can be used in a variety of ways. Some of the uses are to track executives, to determine charges ( highway toll, charge depending on place of visit), to search for stolen items (cars), to study the migratory patterns (animals), to validate legal claims, etc.

## 9.12 Airborne Gravimetry

Kinematic GPS techniques can be used to accurately position airplanes in flight. If the position of the airplane, and more importantly its vertical acceleration and tilt can be monitored with GPS, airborne gravimetric measurements can be corrected for non-gravitational accelerations.

Airborne gravimetric data can be used for scientific research and natural resource exploration.

## 9.13 Other Uses

Some of the important applications of GPS technology are surveying, navigation of missiles, electric power synchronization, agriculture, forestry, census taking, and backpacking emergency systems, and natural resource management.

# 10   Commercial Efforts

Many commercial efforts for using the GPS technology in innovative ways have been announced. A few of the efforts are described briefly here.

## 10.1   Motorola

The Cellular Positioning & Emergency Messaging Unit communicates GPS-determined vehicle position and status. It is 6 inches by 5.5 inches by 2 inches, weighs 16 ounces and has a DB25 cellular interface. Its serial communications are RS-232 with RTS 9600 baud. Motorola has also developed Traxar hand-held navigational computer.

## 10.2   Seiko Communications

Seiko has announced global wireless information services using FM radio. The transmissions include differential GPS data.

## 10.3   Fujitsu

The Car Marty, vehicle multimedia device ($2,640), consists of a main system box, a CD player, and a 5.6-inch color TV monitor (supports TV programs). All the software for the device is provided on CD-ROMs or IC (integrated circuit) cards. The unit provides navigation based on the GPS data.

## 10.4   Sony Mobile Electronics and Etak Inc.

They have introduced a computerized navigation system for automobiles. A car's location shows up on a detailed map. It runs on laptop computers and requires a PCMCIA compliant attachment that processes data from GPS satellites. Drivers can pinpoint their location and track their driving progress in real time. Etak provides software for the maps as well as information on restaurants, hotels, entertainment and shopping in local areas. The system, which will sell for around $2,200, includes a map disc player, a GPS antenna, a five-inch, color LCD display and a wireless remote control. The disc player, which plugs into a cigarette lighter, integrates an eight-channel parallel GPS receiver and a CD-ROM drive.

## 10.5   Penstuff & Trimble Navigation

These two companies have developed GPS standard for PenPoint. By combining GPS technology with pen computers, users can pinpoint their location in terms of latitude, longitude and altitude.

## 10.6 DeTeMobil

They plan to have GPS receivers in all cars in Germany. It will facilitate paying of tolls using smart cards and GSM digital phone.

## 10.7 Monterey Bay Aquarium Research Institute

Their ship-shore WAN includes a remotely operated vehicle node. Here, GPS technology is used for location information.

## 10.8 Ford

Ford is planning the use of GPS based car alarms to locate stolen cars, to track vehicles, and to recover vehicles. They also plan to develop GPS devices for traffic control, navigation, and mapping.

## 10.9 PacTel Cellular Wireless & Wireless Solutions Inc

They are developing vehicle tracking and location devices.

## 10.10 Trimble

Trimble has formed alliances separately with several companies to develop vehicle tracking and location devices. Examples are Trimble & Bell Atlantic and Trimble & IBM.

## 10.11 Avis

Avis is testing GPS navigational aids in rental cars in NYC area.

## 10.12 Toshiba

Toshiba has also developed a portable navigation system.

## 10.13 Tusk Inc

Tusk 386 is an all terrain supertablet pen computer with GPS.

### 10.14 Qualcomm

Developed a GPS data link systems for flight test and training ranges.

# 11 Current Distributed Systems and Networking Applications of GPS

## 11.1 Measure Network Delays

GPS technology can be used to measure the network delays. Wandel & Goltermann Inc. developed a DA-30 Internetwork Analyzer that uses GPS to make latency measurements between Ethernet LANs linked by a wide area network. Boards installed in the local and remote DA-30s lock into the GPS time signal broadcasts. The software then conducts latency trials at various intervals between frame transmissions. Tests are claimed accurate to within 150 microseconds. The connection requires two kits priced at $6,750 each.

## 11.2 SONET Clock Distribution

Synchronous Optical Network (SONET) is the transmission technology used by the phone company to transmit high-speed signals over fiber. The method is synchronous and, therefore, requires that clocks at various nodes be kept closely synchronized. Currently, SONET uses out-of-band clock distribution to keep clocks at various nodes synchronized.

In networks there is a need to accommodate different time delays when multiple bit streams terminate in a single network element such as a long distance message switch. For this purpose, synchronous communications systems relies on accurate frequencies being available.

CCITT Recommendation G.811 specifies that all clocks at network nodes should have a long term frequency departure $\leq 10^{-11}$. Stratum 1 (ST-1) is the highest quality clock for a network. Building Integrated Timing Supply (BITS) is the Bellcore's stand-alone clock-system specification for a timing supply. BITS allows using LORAN/Rubidium ST1 clock systems. In near future, they plan to use GPS frequency reference at each SONET site [41].

To provide a high level of performance, AT&T has developed a primary reference clock (PRC) system using the timing signals from GPS. In PRCs, GPS receivers provide long term timing accuracy and rubidium oscillators provide short-term stability [10]. The GPS timing signal synchronizes a series of master clocks which in turn produce signals that are distributed throughout the network.

# 12 Potential Applications of GPS to Distributed Systems and Networks

In this section we discuss some of possible applications of GPS. Most of these are our ideas that need to be explored further for feasibility and cost effectiveness. The applications using precise timing provided by GPS are discussed first. Position applications are discussed afterwards.

## 12.1 Circuit Switching Using Synchronized Clocks

If all the clocks are synchronized, achieving circuit switching becomes easy. The switching schedule is precomputed and used to switch traffic slots. This is similar to synchronized lights on roads. The motorists do not have to stop if various lights have synchronized clock.

## 12.2 Synchronous Slotted Systems

Slotted systems are less sensitive to distance bandwidth product. Therefore, slotted access is more suitable for high-speed or long-distance networks. That is one reason why DQDB, which was being designed for metropolitan area network was slot based. Same reasoning leads us to use slotted access for very high speed networks. In fact, slotted architectures have been proposed for all-optical, multi-gigabit networks.

The main problem with slotted architectures is that a tight clock synchronization is required. This is now possible with GPS. In fact, in an all-optical ARPA research project, GPS clocks were planned at each switch and end-system.

## 12.3 Clock Synchronization in Distributed System

A distributed system's clocks should have the following properties. At any instant the values of the clocks at any two sites in the distributed should differ by no more than $\Delta$. At any instant the value of a clock at a site should differ from Universal time by no more than $\Delta/2$. The values assumed by any clock should be monotonically increasing. The first property cannot be achieved absolutely, but only with some very high probability.

Good clock synchronization is essential for ordering of events (e.g., FCFS scheduling), consistent update of replicated data, at most once receipt of messages, authentication tickets in some systems (e.g., Kerberos), ensuring atomicity, expiration of privileges, prearranged synchronization, ordering multi-version objects and interpreting data that is a function of time.

Issues to be handled in clock synchronization include the fact that clocks do not always run at the precise rate, sites in a distributed system cannot communicate infinitely often, there are always unpredictable delays in the message delivery, and faulty or malicious sites may provide incorrect or inconsistent time values to other sites.

Clock synchronization can be of two types: internal clock synchronization and external clock synchronization. Some technologies for time service are Network Time Protocol (NTP), OSF Distributed Time service (OSF-DTS) based on Digital's DECdts, Fuzzbal routing protocol, and UNIX 4.3bsd time demon *timed*. Various time sources are described in Section 6.1 on time and frequency alternatives.

GPS clock can be used as a source for external clock synchronization. This technology has the potential to provide time accuracy of up to one nanosecond. Such accurate clock synchronization can help improve performance of distributed systems.

In most architectures, there are some servers that cater to the time needs of their clients. Only a few servers have a time provider (external time source); rest of the servers execute protocols to estimate the time. As GPS clocks become less expensive, more servers can act as primary sources. Use of less accurate GPS clocks to increase the fault tolerance is a possibility. Some of the implications could be less complicated protocols, improved fault tolerance, longer synchronization intervals, and better synchronization.

## 12.4 Database Synchronization

Most database synchronization algorithms (after a failure or disconnected operation) make use of logs with timestamp to decide the order of actions. With GPS synchronized clocks, time stamps have higher reliability (and may have higher resolution).

## 12.5 Connectionless Real-time Communication

When trying to support delay guarantees on IP-like connectionless networks, one problem that comes up often is how to schedule service and meet deadline guarantees. With accurate clock times, a deadline timestamp on the packet can be used to find the next packet to schedule.

Using GPS technology, one can achieve a near global clock at intermediate switches in a network. If all intermediate nodes have accurate clocks, an absolute time based real time communication protocol can be used to meet the deadline requirements. In this protocol each packet can be assigned an absolute time frame and all the intermediate points will schedule the communication to meet the deadline.

Similar approach can be taken for scheduling subtasks of real-time and mission-critical applications. By maintaining a global clock, better guarantees can be given.

## 12.6 One-Way Delay

Currently, one way delay is not measurable because the clocks at two stations are never same or synchronized. Even in the design of ATM networks, the issue of one-way delay came up. The ATM traffic management parameters are different for LAN, MAN, WAN, and GAN (global area network). With GPS synchronized clocks at the source and destination, the exact one-way delay between source and destination and to every switch can be easily measured with a single timestamp.

## 12.7 Delay Based Routing

Many networks including internet use link delays for routing. However, accurate measurement of delays is difficult and therefore, approximate or round-trip delay is used. With GPS, exact one-way delay can be used for routing.

## 12.8 Time to Live

The "time to live" field in packets is used to remove old packets from networks. Currently, the time-to-live field is decremented by 500 ms regardless of actual delay. With GPS synchronized clock, time-to-live could be the actual time.

## 12.9 Diagnostics/Maintenance of system clocks

A GPS frequency calibrator can be used to periodically check crystals in various equipment.

## 12.10 Resource Location

With the help of digitized maps and information about the environment (over radio links or disks) Mobile units can locate and use the nearest resources.

The GPS derived position can also be used to locate replicated resources, e.g., the nearest printer or fileserver in a mobile environment (even in the event of failure of some resources).

A location dependent server migration approach to provide location independent computing is possible with the use of GPS derived parameters.

## 12.11 Location Adaptive Protocols

Networking today is location transparent. All service scheduling decisions are made without regard to the location. There are many applications where knowing the location will help us adapt a different service strategy. Examples of such location-adaptive protocols are:

### 12.11.1 Home vs Office vs Car

The location decides the physical medium (wire, ISDN, modem, cellular, or radio) available to the user and hence the available bandwidth, cost, and error characteristics. My laptop has a high bandwidth connection when at the office. It has 28.8 kbps connection from home. But from the car, it has an expensive and error-prone 14.4 kbps connection. GPS provided location information can be used to make mobile computing decisions including which files to fetch and at what speed.

### 12.11.2 Electronic Fence

Most companies have physical walls outside of which paper information is not allowed to pass. With GPS the same type of protection can be applied to electronic information. The information is usable only as long as the mobile computer is within the corporate boundary.

## 12.12 Handoffs in Wireless Networks

In cellular architectures, handoffs take place when the mobile communicating unit moves into another cell (inter-cell handoff) or when strong interference on the channel prevents communication over it (intra-cell handoff). In the former case, the connection has to be switched to a new base station and in the later case it can be handled by switching to a different channel. Handoffs can be performed in a centralized fashion (one central switch makes the decision) or decentralized fashion (mobile units make the decision).

In handoffs, the signal strength from various base stations is measured and a decision regarding a handoff is made using the measurements obtained. Handoff's can be made based on the GPS derived position. Using the information about cell boundaries and the position, a decentralized handoff can be initiated.

By this approach, passive listening to beacons can be avoided; complicated handoff protocols can be simplified; intra-cell hand-overs can be detected easily; many unnecessary handoffs can be avoided; and intelligent and optimistic handoff techniques can be employed.

## 12.13 Prescheduled Uninterrupted Handoffs

One problem with signal strength based handoffs is that it is difficult to predict the future. On the other hand, given location, velocity, and direction information, future location can be easily predicted. Signal strength based handoffs may cause some interruption in service as the packets sent to the previous base have to be forwarded to the new base. With GPS, the mobile units position, velocity, and direction information can be used to forecast the next base and prenegotiate the hand-over with all parties.

## 12.14 Adaptive Transmission Power Control Algorithm

Energy efficient computing and communication is a key factor to the success of mobile computing. The expected increase in the battery lifetime is only 20% over the next 10 years and mobile hosts will have limited available power. Adaptive power control algorithms try to optimize the power required for transmission. If the base is nearby, less power is used compared to when the base is far away. Such algorithms can use GPS derived position and direction information.

These algorithms offer two fold advantage in cellular architectures by using less power and by allowing frequency reuse in the same cell. These algorithms can also be developed for mobile communications in the absence of base stations – for peer-to-peer communication, for example, in battlefield communications.

## 12.15 Directional Antennas

Currently mobile units transmit energy equally in all directions. Only part of this energy – the part that is transmitted towards the receiver – is useful. The rest is lost in space. With GPS provided position, the antennas can be directional. This will help reduce the energy required, battery drainage, and, hence the weight of the units.

Directional antennas are particularly helpful for long-distance wireless communication, e.g., satellite communication.

Directional antennas also allow better packing density – more users for the same space.

With directional antennas, the mobile units can talk to the least-busy base unit even if it is not closest unit.

This also helps in providing the minimum radiated RF pattern for covert communications.

## 12.16 Temporary Cell Partitioning for Congestion Avoidance

In cellular architectures for mobile communications cell splitting is a process of dividing a current cell to form new cells. This increases the reuse of spectrum and helps in reducing the congestion. This process requires prior preparation and usually a permanent change. Using position based handoffs, digitized cell layout information and directional antennas, cell splitting may be possible on a temporary basis. The above approach can also be used in case of base station failures.

## 12.17 Peer-to-peer Routing with Limited Range Receivers

Most of the civilian wireless communication today uses pre-established infrastructure such as base units. In military environment (for example, operation desert storm), usually there are no pre-existing infrastructures. In such environments, it is better to use peer-to-peer communication where signals from one soldier are transmitted and forwarded by another. Each mobile unit acts as an end-system as well as a router. Each unit is aware of the GPS position of other units and uses this information to find the shortest route to the destination.

To be effective in these networks, optimum networking strategy must be adaptive with respect to the possible rates of platform movements. Approaches that use RF link quality as the primary controlling parameters for the network routing decisions may not be adequate for highly mobile platforms. GPS provided position, heading, velocity, as well as, digital terrain topology information can be used in determining the optimum routing for packet data.

### 12.18 Email Delivery Based on Geographic Location

In wired networks, messages are routed based on an address. The addresses have no relation to the physical location. In some applications, one may want to multicast/anycast to a particular geographic location. For example, one may want to address a message to all police cars near Stanford university on route 101. Such location addressing is feasible with GPS equipped receivers.

In particular one could build a "Cellular data service" with towers all over the country. The GPS coordinates of the mobile unit will help in routing.

### 12.19 Distributed Robot Control and Navigation

GPS technology can be applied to robots for tracking and control purposes. Location dependent actions can also be performed. Intelligent robots can use position information usefully. Distributed robots (a collection of robots for some task) can be controlled better with GPS. Unmanned vehicles can navigate effectively. Position based controls instructions can be executed by the robots (location dependent actions).

### 12.20 Equipment Location Marking for Maintenance Crew

Service requesters (mobile or stationary) provide GPS location. Maintenance crew carry GPS to locate the equipment.

## 13 Current Limitations of GPS

The use of GPS technology poses some problems depending on the application, the type and size of the receiver being used and the type of access being allowed (SPS/PPS). Some of the problems are Selective Availability which degrades the achievable accuracies, low signal strength, weight and size of accurate GPS receivers, power consumption for mobile applications, temporary outage of the receiver as the receiver passes under obstructions, etc.

Due to the uncertainty of availability of GPS, GPS clocks can be used only for optimization not for operation. The systems should continue to work without the GPS. This is like cache memories. All systems work with or without cache. They work better with cache.

# 14  GPS Products and Services

The market for GPS devices was estimated to be $200 million (sold) in 1992 and could reach $500M in 1995. These devices include receivers, modems, simulators, range finders, robotic systems, and navigation modules.

## 14.1  GPS Receivers and Stations

### 14.1.1  General Purpose Handheld Receivers

Magellan NAV 1000, Magellan NAV 5000, Motorola Commando, Motorola Traxar, Rockwell Trooper, Trimble Pathfinder.

### 14.1.2  Land Vehicle Receivers

Rockwell PathMaster.

### 14.1.3  Mapping Receivers

Corvallis Mictrotechnology MC-GPS, Motorola LGT-1000 Lightweight GPS/GIS Terminal, Sokkia Spectrum.

### 14.1.4  Aircraft Receivers

Ashtech ALTAIR AV-12.

### 14.1.5  Miscellaneous Receivers

GEC Plessey, Novatel.

### 14.1.6  OEM Receivers

Ashtech Sensor II, Garmin GPS 10, Garmin MultiTrac 8, Garmin PhaseTrac 12, Garmin TracPak, JRC GPS Receiver Core, Koden GPS-1A, Leica GPS Engine, Magellan AIV-10, Motorola Oncore, Rockwell Navcore, Novatel (OEM Kit family), Rockwell NavCore V GPS Sensor, Trimble SVeeSix.

### 14.1.7 PC Card (PCMCIA) Receivers

Rockwell Navcard, Trimble GPS card, Trimble GPS Gold card.

### 14.1.8 Shipboard Receivers

Adroit Systems TRIADS, Leica MX100 Navigator, Motorola Peregrine DGPS Navigation System.

### 14.1.9 Spacecraft Receivers

Motorola Monarch

### 14.1.10 Surveying Receivers

Allen Osborne Associates TurboRogue, Ashtech Dimension 12, Ashtech Z-12, Del Norte Model 586 DMU/1008 GPS, Geotronics Geotracer 2000, Leica SR 261, Leica System 20, Trimble 4000 SE, Trimble 4000SSE, Trimble 4000 ST, Trimble 4000 SST.

### 14.1.11 Timing Receivers

Allen Osborne Associates TTR-6 Time & Frequency Receiver, Allen Osborne Associates TTR-4P Time & Frequency Receiver, Austron Model 2200 GPS Receiver, Ball GPS-RR GPS Referenced Time & Frequency Std, Bancomm (Family of receivers), Electronik GmbH HOPF, TRAK Model 8810 GPS Station Clock, Stellar Model 100 GPS Clock.

## 14.2 Modems

CSI MBX1 USCG Beacon Receiver, Racal DeltaFix LR MF/HF Transmitter-Receiver, NavSymm DR5-96S GPS Differential Transceiver, and Pacific Crest DDR-96 UHF Radio Modem.

## 14.3 Satellite Simulators

Northern Telecom STR2760 Multi Channel GPS Simulator, Northern Telecom STR2770 Single Channel GPS Simulator, Stanford Telecom Model 7201 Single Channel C/A Generator.

## 14.4 Laser Range Finders

Laser Atlanta ProSurvey 1000, Laser Technology Criterion, Leica.

## 14.5 Robotic Total Stations

Geotronics Geodimeter System 5000

## 14.6 Navigation Modules

KVH Industries C-100 Compass Engine, Precision Navigation TCM1 Electronic Compass Module.

# 15 Details of A Few Selected Products

## 15.1 Magellan System M2

Real time positioning using base station generated RTCM-104 DGPS corrections. Consists of M2a base station (DGPS message generator $9,740), M2b remote receiver($2,795) and M2c wireless modem ($23,780). Provides differential accuracy of 3-10 meters.

## 15.2 Real time DGPS systems

Using cost guard beacons or FM subcarrier corrections, these provide real time accuracies of 3-10 meters. Magellan Trailblazer XL ($1045), Trimble ScoutMaster($1,575), Magellan 10 channel Sensor($1,191), Magellan 5000 DLX ( $1,395) and Magellan FieldPRO V ($2,091).

## 15.3 Mobile GPS Card

Type II PCMCIA GPS sensor by Trimble ($795). It has 3 channels tracking up to 8 satellites. Achievable accuracy is 100 meters. It has an acquisition time of less than 30 seconds and re-acquisition rate of 2-3 seconds.

## 15.4 Mobile GPS Gold Card

Differential-ready Type II PCMCIA GPS sensor and development system from Trimble ($1,595). Provides 2-5 meters accuracy in real-time.

## 15.5 Mobile GPS Intelligent Sensor 100

By Trimble Navigation ($395 estimated). Trimble Navigation's low-end sensor.

## 15.6 NavCard PCMCIA GPS sensor

Rockwell Type II PCMCIA GPS receiver. Time to fix: 30 seconds from warm start and 15 min cold start, optional differential capability. Accuracies of 13.7 cm × 5.4 cm.

## 15.7 Stellar Model 100 GPS clock

100 MHz standard output with an accuracy of $5 \times 10^{-12}$, corrected 1 pps output with less than 1 nanosecond jitter ($2,795).

## 15.8 Rockwell Portable MicroTracker GPS receiver

MicroTracker operates with inexpensive, passive antenna in most applications. It can track as many as nine satellites simultaneously, and has a time-to-first fix of 20 to 30 seconds (from a warm start).

## 15.9 Mobile Computing Kit

It includes pen-based TelePad, Proxim's RangeLAN, cellular phone, Trimble GPS, FotoMan Plus camera, ScanMan, AudioMan ($7,299).

## 15.10 GPS Software Applications

A lot of commercial activity is being undertaken on the software front. Details of a few of the softwares are listed below.

1. **GPS for windows ($1,995):** By Peacock Systems. It gives location, speed, direction. It passes position via DDE. It supports access to the data in real time. The package includes a GPS receiver and an antenna.

2. **GPS Signal Simulation software:** Accord Software and Systems. ($495). Software package for understanding the characteristics of the GPS satellite signal and its processing. Platform to experiment with the various modules of that constitute the signal processing section of a GPS receiver.

3. **City Streets for Windows:** Software alone costs $99.95 from Road Scholar software, 800-426-7623. Optional $399.95 Marco Polo PCMCIA card adds GPS support. It provides interactive location information Can substitute Trimble card for $599.95.

4. **Streets on a Disk:** By Kylnas Engineering ($225+$95 per county) has the digitized maps of streets on disks. This can be used with other GPS products.

5. GPSez and GPSpac for Windows: ($1,290) by General Engineering and Systems S.A.

6. MapInfo for Windows 3.0: MapInfo Corp.

7. Atlas GIS for Windows 2.0: By Strategic Mapping Inc.

8. GISPlus for PC: By Caliper Corp.

9. Zagat-Axxis CityGuide: Zagat restaurent guide by Axxis Software.

10. Maptech Professional Marine Chart S/W: ($1,290) by Resolution Mapping Inc.

11. GPS MapKit XV: By DeLorme Mapping. It links GPS to maps.

12. Map'n'GO: ($50) 3CS Software.

13. NCompass 3.0 for Windows: - real time GPS, 312-271-1020

## 15.11 Scientific GPS Software Suites

A list of major scientific software suites (platforms) are given below. For further details regarding these, please refer [5]. BAHN/GPSOBS (IBM MVS/XA,UNIX), BERNESE (UNIX,PC DOS,VMS), CGPS22 ( UNIX), DIPOP ( UNIX), EPOS.P.V3 (CONVEX,SUN,CRAY,IBM,CYBER), GAMIT/GLOBK ( UNIX), GAS (IBM-PC,UNIX,VME), GEOSAT ( UNIX ), GIPSY/OASIS ( UNIX, HP9000/7000), MSOP (VP-2600 and NWT), PAGE3 ( HP9000, DOS) and TEX-GAP/MSODP (Cray Y-MP).

# 16 DGPS Correction Signal Services

Some of the efforts by various organizations to provide DGPS correction signals are described below.

## 16.1 Differential Corrections Inc.

The DCI system transmits DGPS correction signals over FM broadcast station subcarriers using Radio Data System (RDS) protocols. The system requires the operator to purchase a receiver and sign a service contract. Receiver costs around $300. Several levels of services are available.

## 16.2 John Chance & Associates

The Omnistar system supplies the user with DGPS corrections in RTCM 104 format. The user receives correction signals from a commercial geostationary satellite. The system is fed by 10 differential stations. The system requires the purchase a satellite receiver and signing a service contract. It costs around $6,000 for the service and $5,000 for the receiver. Through this system, corrections are available across the entire United States and much of Canada and Mexico.

## 16.3 Accqpoint

The Accqpoint system transmits DGPS correction signals in RTCM-104 format. Transmission is over FM subcarrier using Radio Data System (RDS) protocols. User needs to purchase a receiver (around $300) and sign a service contract.

## 16.4 Cue Network Corp. & Differential Corrections Inc

Nationwide differential GPS network of local FM radio stations will provide position information, accurate to within five meters. Will provide information for surveying, vehicle tracking and geographic information systems (GIS).

# 17  Addresses of GPS Equipment Manufacturers

- **Allen Osborne Associates**, 756 Lakefield Road, #J, Westlake Village, CA 91361, Phone: 805-495-8420, Fax: 805-373-6067.

- **Ashtech**, 1170 Kifer Road, Sunnyvale, CA 94086, Phone: 408-524-1400 or 1-800-229-2400, Fax: 408-524-1500.

- **Auspace Limited**, 50 Hoskins Street, Mitchell ACT 2911, Australia, Phone: (06)242-2611, Fax: (06)241-6664.

- **Best-Fit Computing**, PO Box 35103, Tucson, AZ 85740, Phone: 602-544-2432.

- **Collins Avionics & Communications**, Dept. 153-250, 350 Collins Rd. NE, Cedar Rapids, Iowa 52498, US Phone:: 800-321-2223, World Phone:: 319-395-5100, Fax: 319-395-4777.

- **Del Norte Technology, Inc.**, 1100 Pamela Drive, PO Box 696, Euless, TX 76039, Phone: 817-267-3541, Fax: 817-354-5762.

- **FTS Austron Inc.**, PO Box 14766, Austin, TX 78761-4766, Phone: 512-251-2313, Fax: 512-251-9685.

- **Garmin International**, 9875 Widmer Road, Lenexa KS 66215, Phone: 913-599-1515 or 800-800-1020, Fax: 913-599-2103.

- **Garmin/Europe Ltd**, Robert House, Station Approach, Romsey, Hampshire S051 8DU, Phone: +44-794-519944.

- **GEC-Plessey**, Martin Road, West Leigh,, Havant, Hants PO9 5DH, United Kingdom., Fax: 44 (705)493604.

- **Gnostech Inc.**, 650 Louis Drive, Suite 190, Warminster, PA 18974, Phone: 215-443-8660.

- **Intermetrics California Division**, 5312 Bolsa Avenue, Huntington Beach, CA 92649, Phone: 714-891-4631.

- **Japan Radio Co., Ltd.**, Akasaka Twin Tower (Main), 17-22, Akasaka 2-chome, Minato-ku, Tokyo, 107 Japan, Phone: 03-584-8844, Fax: 03-584-8878.

- **JRC(UK)Ltd.**, Rm. No. 52/54, Ground Floor,, Temple Chambers, Temple Ave., London E.C. 4, England, Phone:: 71-353-7960, fax: 71-353-3321.

- **Japan Radio Co., Ltd.**, 430 Park Ave., New York, NY. 10022, USA.

- **John E. Chance & Associates, Inc.**, 200 Dulles Drive, Lafayette, LA 70506, Phone: 318-237-1300, Fax: 318-237-0011.

- **Kinemetrics/Truetime**, 3243 Santa Rosa Avenue, Santa Rosa, CA 95407, Phone: 707-528-1230, Fax: 707-527-6640.

- **Leica, Inc.**, 23860 Hawthorne Blvd. CH-9435 Heerbrugg, Switzerland, Torrance, Ca. 90505, USA, Phone:: 310-791-6116 +41(071)703 131, fax: 310-791-6108 +41(071)703 998.

- **Leica**, 45 Epping Rd,, North Ryde NSW 2113, Australia, Phone: (02)888 7122, Fax: (02)888 7526.

- **Magellan Systems Corporation**, 960 Overland Court, San Dimas CA 91773, Phone: 909-394-5000 or 800-669-4477, Fax: 909-394-7050.

- **Magnavox Advanced Products and Systems Co.**, 2829 Maricopa Street, Torrance, CA 90503, Phone: 310-618-1200, Fax: 310-618-7001.

- **Micrologic**, 9610 DeSoto Ave., Chatsworth CA 91311, Phone: 818-998-1216, Fax: 818-709-3658.

- **Motorola Military and Aerospace Electronics Inc.**, 8201 E. McDowell Road, PO Box 1417, Scottsdale, AZ 85252, Phone: 800-544-3934, Fax: 602-441-7391.

- **NAVSTAR Electronics Inc.**, 1500 North Washington Blvd., Sarasota, FL 34236, Phone: 813-366-6338, Fax: 813-366-9335.

- **Navstar Ltd.**, Royal Oak Way, Daventry, Northants, NN11 5PJ, England, Phone: 0327 79066, Fax: 0327 72491 and 71116.

- **Novatel Communications Ltd.**, 6732 - 8th Street NE, Calgary, Alberta, Canada T2E 8M4, Phone: 403-295-4500, Fax: 403-295-0230, GPS Hot Line 403-295-4900, Fax: 403-295-4901.

- **Rockwell International**, 4311 Jamboree Road, PO Box C, Newport Beach, CA 92658-8902, Phone: 800-436-9988, Fax: 818-365-1876.

- **Rockwell International, Ltd.**, Central House, 3, Lampton Road, Hounslow, Middlesex, TW3 1HY England, Phone: (44-81)751-6779, Fax: (44-81)570-0758.

- **Rockwell International (Taiwan)**, Digital Communications, Division,, Room 2808, International Trade Building,, 333 Keelung Road, Section 1, Taipei, Taiwan 10548, ROC, Phone: (886-2)720-0282, Fax: (886-2)757-6760.

- **Rockwell International**, Digital Communications Division, 3 Thomas Holt Drive, PO Box 165, North Ryde NSW 2113, Australia, Phone: (02)805 5555.

- **Sokkia**, Rydalmere Metro Centre, Unit 29 38-46 South Street, Rydalmere NSW 2116, Australia, Phone: (02)638 0055, Fax: (02)638 3933.

- **Sealevel Systems Inc.**, PO Box 1808, Easley, SC 29641, Phone: 803-855-1581.

- **Sextant Avionique**, Navigation Systems Division, 25, rue Jules Vedrines, 26027 Valence Cedex, France, Phone: 33-75 79 8511, Fax: 33-75 55 2250.

- **Sokkisha Co. Ltd.**, 1-1, Tomigaya 1-Chome, Shibuya-Ku, Tokyo, 151 Japan, Phone: 03-465-5211, Fax: 03-465-5203.

- **Stanford Telecom**, 2421 Mission College Boulevard, Santa Clara, CA 95054, Phone: 408-980-5684, Fax: 408-980-1066.

- **Starlink Incorporated**, 1321 Rutherford Lane #180, Austin, Texas 78753, Phone: 512-832-1331, Fax: 512-832-7857, Contact: David Fowler, e-mail: dfowler@bga.com.

- **STC Government Systems Inc.**, 591 Camino de la Reina, Suite 428, San Diego, CA 92108, Phone: 619-295-5182.

- **STC Navigation Systems**, London Road, Harlow, Essex CM17 9NA, UK, Phone: +44 (0)0279 429531.

- **Trimble Navigation**, 645 North Mary Avenue, PO Box 3642, Sunnyvale, CA 94088-3642, Phone: 1-800-827-8000 or 408-481-8000, Fax: 408-481-6640.

- **Trimble Navigation Europe Ltd.**, Trimble House, Meridian Office Park, Osborn Way, Hook, Hampshire RG27 9HX, England, UK, Phone: +44 256-760150, Fax: +44 256-760148.

- **Trimble Navigation France**, ZAC du Moulin, 9 rue de l'Arpajannais, 91160 Saulx les Charteux, France, Phone: +33 1 64-54-83-90, Fax: +33 1 64-34-49-73.

- **Trimble New Zealand**, 76 Chester Street East, PO 13-547, Armagh, Christchurch, New Zealand, Phone: (64)3-371-3400, Fax: (64)3-371-3417.

- **Van Martin Systems, Inc.**, PO Box 2203, Rockville, MD 10847, Phone: 301-468-2095, Fax: 301-770-6555.

# 18  Appendix: Acronyms

| | |
|---|---|
| AE | Antenna Electronics |
| AHRS | Attitude and Heading Reference System |
| A/J | Anti-Jamming |
| AOC | Auxiliary Output Chip |
| A-S | Anti-Spoofing |
| ASIC | Application Specific Integrated Circuit |
| ATE | Automatic Test Equipment |
| BIH | Bureau International de L'Heure |
| BIPM | International Bureau of Weights and Measures |
| BIT | Built-In-Test |
| C/A-code | Coarse/Acquisition-Code |
| CADC | Central Air Data Computer |
| CDU | Control Display Unit |
| CEP | Circular Error Probable |
| C/No | Carrier to Noise Ratio |
| CRPA | Controlled Radiation Pattern Antenna |
| CSOC | Consolidated Space Operations Center |
| CW | Continuous Wave |
| DAC | Digital to Analog Converter |
| DGPS | Differential GPS |
| D-Level | Depot Level |
| DLM | Data Loader Module |
| DLR | Data Loader Receptacle |
| DLS | Data Loader System |
| DMA | Defense Mapping Agency |
| DoD | Department of Defense |
| DOP | Dilution of Precision |
| DRMS | Distance Root Mean Square |
| DRS | Dead Reckoning System |
| DT&E | Development Test and Evaluation |
| ECEF | Earth-Centered-Earth-Fixed |
| EDM | Electronic Distance Measurement |
| EFIS | Electronic Flight Instrument System |
| EM | Electro Magnetic |
| EMCON | Emission Control |
| ESGN | Electrically Suspended Gyro Navigator |
| FAA | Federal Aviation Administration |
| FMS | Foreign Military Sales |
| FOM | Figure Of Merit |
| FRPA | Fixed Radiation Pattern Antenna |
| FRPA-GP | FRPA Ground Plane |
| GDOP | Geometric Dilution of Precision |
| GMT | Greenwich Mean Time |
| GPS | Global Positioning System |
| HDOP | Horizontal Dilution of Precision |
| HOW | Hand Over Word |
| HSI | Horizontal Situation Indicator |
| HV | Host Vehicle |
| HQ USAF | Headquarters US Air Force |
| ICD | Interface Control Document |
| ICS | Initial Control System |
| IF | Intermediate Frequency |
| IFF | Identification Friend or Foe |
| I-Level | Intermediate Level |
| ILS | Instrument Landing System |
| INS | Inertial Navigation System |
| ION | Institute of Navigation |
| IOT&E | Initial Operational Test and Evaluation |
| IP | Instrumentation Port |
| ITS | Intermediate Level Test Set |
| JPO | Joint Program Office |
| J/S | Jamming to Signal Ration |
| JTIDS | Joint Tactical Information Distribution System |
| L1 | GPS primary frequency, 1575.42 MHz |
| L2 | GPS secondary frequency, 1227.6 MHz |
| LEP | Linear Error Probable |
| LRIP | Low Rate Initial Production |
| LRU | Line Replaceable Unit |
| LO | Local Oscillator |
| mB | Millibar |
| MCS | Master Control Station |
| MCT | Mean Corrective Maintenance Time |
| MHz | Megahertz |
| MLV | Medium Launch Vehicle |
| MmaxCT | Maximum Corrective Maintenance Time |
| MOU | Memorandum of Understanding |
| M/S | Metres per Second |
| MSL | Mean Sea Level |
| MTBF | Mean Time Between Failure |
| MTBM | Mean Time Between Maintenance |
| NAV-msg | Navigation Message |
| NOSC | Naval Ocean Systems Center |
| NRL | Naval Research Laboratory |
| NS | Nanosecond |
| NSA | National Security Agency |
| NTDS | Navy Tactical Data System |
| NTS | Navigation Technology Satellite |
| OBS | Omni Bearing Select |
| OCS | Operational Control System |
| O-Level | Organization Level |
| OTHT | Over The Horizon Targeting |
| P-Code | Precise Code |
| PDOP | Position Dilution of Precision |
| PLSS | Precision Location Strike System |
| P I | Pre Planned Product Improvement |
| PPS | Pulse Per Second |
| PPM | Pulse Per Minute |
| PPM | Parts Per Million |
| PPS | Precise Positioning Service |
| PPS-SM | PPS Security Module |
| PRN | Pseudo Random Noise |
| PTTI | Precise Time and Time Interval |
| PVT | Position Velocity and Time |
| RAM | Reliability and Maintainability |
| RCVR | Receiver |
| RF | Radio Frequency |
| RMS | Root Mean Square |
| RNAV | Area Navigation |
| RSS | Root Sum Square |
| RT | Remote Terminal |
| RTCA | Radio Technical Commission for Aeronautics |
| RTCM | Ratio Technical Commission for Maritime Services |
| S/A | Selective Availability |
| SAMSO | Space and Missile Systems Organization |
| SBB | Smart Buffer Box |
| SC | Special Committee |
| SEP | Spherical Error Probable |
| SI | International System of Units |
| SIL | System Integration Laboratory |
| SINS | Shipborne INS |
| SPS | Standard Positioning Service |
| SRU | Shop Replacable Unit |
| STDCDU | Standard CDU |
| TACAN | Tactical Air Navigation |
| TAI | International Atomic Time |
| TDOP | Time Dilution of Precision |
| TFOM | Time Figure Of Merit |
| TTFF | Time to First Fix |
| UE | User Equipment |
| UERE | User Equivalent Range Error |
| UHF | Ultra High Frequency |
| USA | United States of America |
| USNO | US Naval Observatory |
| UT | Universal Time |
| UTC | Universal Time Coordinated |
| VDOP | Vertical Dilution of Precision |
| VHSIC | Very High Speed Integrated Circuit |
| VLSIC | Very Large Scale Integrated Circuit |
| VOR | Very High Frequency (VHF) Omnidirectional Range |
| WGS-84 | World Geodetic System - 1984 |
| YPG | Yuma Proving Ground |